\documentclass[sigconf]{acmart}

\usepackage{booktabs} 
\usepackage[colorinlistoftodos]{todonotes}

\usepackage{csquotes}

\usepackage{epsfig}
\usepackage{color}

\usepackage{balance}  
\usepackage{algorithmic}
\usepackage{algorithm}
\usepackage{multirow}
\usepackage{graphicx}
\usepackage{amsmath}
\usepackage{amssymb}
\usepackage{amsfonts}
\usepackage{xspace}
\usepackage{url}

\newcommand{\IGNORE}[1]{}

\newcommand{\commentedtext}[1]{}

\graphicspath{{./figures/}}

\begin{document}

\copyrightyear{2017}
\acmYear{2017}
\setcopyright{acmcopyright}
\acmConference{SIGIR '17}{August 07-11, 2017}{Shinjuku, Tokyo, Japan}\acmPrice{15.00}\acmDOI{10.1145/3077136.3080695}
\acmISBN{978-1-4503-5022-8/17/08}

\title{Predicting Session Length in Media Streaming}


\author{Theodore Vasiloudis}
\authornote{Part of this work was performed during an internship at Pandora Media Inc.}
\affiliation{
	\institution{RISE SICS}
	\streetaddress{Stockholm, Sweden}}
\email{tvas@sics.se}

\author{Hossein Vahabi}
\affiliation{
	\institution{Pandora Media Inc.}
	\streetaddress{Oakland, USA}}
\email{puya@pandora.com}

\author{Ross Kravitz}
\affiliation{
	\institution{Pandora Media Inc.}
	\streetaddress{Oakland, USA}}
\email{rkravitz@pandora.com}

\author{Valery Rashkov}
\affiliation{
	\institution{Pandora Media Inc.}
	\streetaddress{Oakland, USA}}
\email{vrashkov@pandora.com}

\renewcommand{\shortauthors}{T. Vasiloudis et al.}

\begin{abstract}
	Session length is a very important aspect
	in determining a user's satisfaction with a media streaming service. Being able
	to predict how long a session will last can be of great use
	for various downstream tasks, such as recommendations and ad scheduling.
	Most of the related literature on user interaction duration has focused on dwell time for websites,
	usually in the context of approximating post-click satisfaction
	either in search results, or display ads.
	
	In this work we present the first analysis of session length in a
	mobile-focused online service, using a real world data-set from a major music streaming service.
	We use survival analysis techniques to show that the characteristics of the length distributions
	can differ significantly between users, and use gradient boosted trees with appropriate objectives
	to predict the length of a session using only information available at its
	beginning.
	Our evaluation on real world data illustrates that our proposed technique outperforms the considered baseline.

\end{abstract}

\keywords{User Behavior; Survival Analysis; Dwell Time; Session Length}
\maketitle

\section{Introduction}
\label{sec:introduction}
%

More and more people these days use online services to consume media.
Whether they are consuming videos or music, users start an interaction
with a service, consume a number of items, and after some time end their interaction.
We refer to this complete interaction as a \textit{user session}, and the time it takes from
start to finish as the \textit{length} of the user session. 

Being able to predict the length of a session is important, because it allows the service
provider to optimize the user experience along with its business goals.
In terms of user experience, session length can be a useful signal for a
recommendation engine. For a music streaming service, where users
typically consume multiple items in a single session, the recommendation system
can be tuned to be more exploratory or exploitative, based on the expected length
of the session. On the business side, services often need to present ads to users
to generate revenue, but there is a limit on how many ads a user will tolerate
within one session \cite{goldstein2013ads}.
The length of a session provides a vital data point for the trade-off between user satisfaction and
generated revenue.
By having an estimate of the length of a session early on, ads can be rescheduled
so that the revenue target (i.e. number of ads presented) is maintained while
minimizing the annoyance to users.

The length of user sessions can be hard to predict, because of
two main factors: First, there exist a number of extraneous
parameters that can affect session length, that are difficult to model
based only on data that are available to an online service.
Sessions can start and end for any number of reasons: users entering/exiting the subway, 
arriving at home or work etc.
Second, user interactions commonly exhibit long-tail distributions; see for example
dwell time studies \cite{liu2010weibull, Kim2014satisfaction},
phone call duration \cite{seshadri2008mobile} and Section \ref{sec:weibull} of this study.
This, in combination with
the lack of predictive features makes it harder
to correctly place the probability mass of predictive models.
In this work we mitigate this issue by using an appropriate objective function for
our model.

Most of the related research has focused on website visits, modeling the time spent on a clicked result (\textit{dwell time}) after a 
search \cite{Kim2014satisfaction, borisov2016catm} or an ad click \cite{Barbieri2016RSFclick, lalmas2015gemini}. This type
of interaction is very different from the way users consume items on a media streaming service.
In a web search or ad click scenario, users enter a query or click on an ad, check the result
and leave the website, in a ``screen-and-glean'' behavior \cite{liu2010weibull}.
In a media streaming service, users may interact very little with the 
platform, but have very long sessions (``lean-back'' behavior),
or have exploratory sessions, constantly revising their selection until settling down
or abandoning the session.
We show in Section \ref{subsec:weibull-analysis}
that 44\% of the users exhibit ``negative-aging'' length distributions, i.e. sessions
that become less likely to end as they grow longer.

To summarize, the key contributions of this study are the following:

\begin{itemize}
    \item We provide an analysis of user session length in an online media streaming service, using the Weibull distribution in Section \ref{sec:weibull}.
    \item We develop a predictive model for session length using contextual and user-based features
    with appropriate objective functions in Section \ref{sec:prediction} and present experimental
    results in Section \ref{sec:experiments}.
\end{itemize}

\section{Related Work}
\label{sec:related}
%

Survival analysis and prediction of dwell time has come into focus recently,
with many studies using it as a proxy of user satisfaction in
search and ad click scenarios. One of the first studies using
the Weibull distribution analyzed the
dwell time of users
visiting web sites after performing a search \cite{liu2010weibull}. The study indicated a strong
``negative aging'' effect for websites visited after a search
, i.e. the probability of abandoning a page decreased with increasing dwell time (see Sec. \ref{subsec:weibull-review}).

The distribution of dwell time can also be used to measure and improve the
satisfaction of users with search results \cite{Kim2014satisfaction}.
In this study, the authors segment the results according
to attributes like ``readability level'' and model the distribution
of dwell time for each segment. Their findings include that the
dwell time distributions for satisfied and dissatisfied clicks differ,
and that it is possible to use characteristics of these distributions
to better predict satisfied clicks.

Finally, predicted dwell time has also been used to improve the ranking of
ads \cite{Barbieri2016RSFclick, lalmas2015gemini}.
The predicted dwell time is incorporated
into an ``ad-quality'' model, which may include other aspects
of an ad and a user, summarized as the probability of a user clicking on
an ad. The authors develop predictive models and use features about the
user and ad landing page to estimate the dwell time and bounce rate. They perform an evaluation
on historical data and online experiments to measure the effect
that using the ad quality model has on user engagement.

\section{Weibull analysis of session length}
\label{sec:weibull}
%

In this section we perform an analysis of the session length distribution
for users in our sample. We provide a brief introduction into Weibull analysis
then move on to the results and discussion.

\subsection{Weibull Distribution Review}

\label{subsec:weibull-review}

The Weibull distribution is attractive for
survival analysis because it allows us to model different kinds of failure rates,
when the probability of failure changes over time. The probability
density function (PDF) of the distribution is:

\begin{equation}
    \label{eq:weibull-pdf}
    f(t) = \frac{k}{\lambda}\left( \frac{t}{\lambda}\right)^{k-1}e^{-(t/\lambda)^k}, t \geq 0
\end{equation}

\begin{figure}
    \centering
    \includegraphics[width=0.47\textwidth]{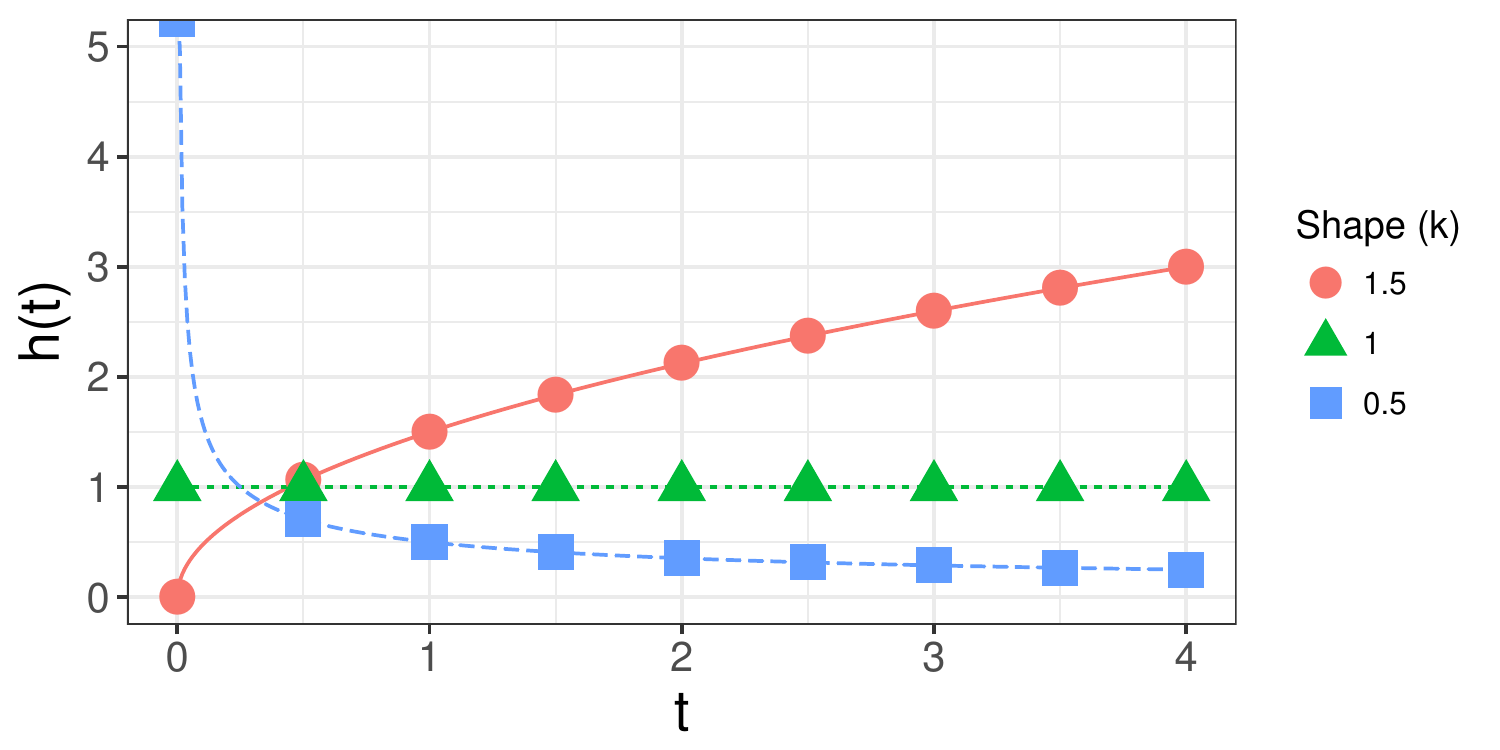}
    \caption{The failure rate of the Weibull distribution for different values of the shape parameter, $k$. We set $\lambda = 1$.}
    \label{fig:weibull-failure-rate}
\end{figure}

The distribution has two parameters, $k$ and $\lambda$, which correspond to the \textit{shape}
and \textit{scale} of the distribution. The shape, $k$, determines how the elapsed
time affects the rate of failure. The scale, $\lambda$, affects the spread of
the distribution: the larger it is, the more spread out the distribution becomes.

The effect of $k$ can be better illustrated by the \textit{hazard rate} (or hazard function) which gives us the
failure rate of an item that has survived up to time $t$.
For the Weibull distribution it is given by:

\begin{equation}
    \label{eq:weibull-failure-rate}
    h(t) = \frac{k}{\lambda}\left( \frac{t}{\lambda}\right)^{(k - 1)}
\end{equation}

The effect of $k$ is illustrated in Figure \ref{fig:weibull-failure-rate}.
For values $0 < k < 1$ the hazard rate decreases as time increases. This behavior is
often described as ``negative aging'' or ``infant mortality'' failures, where defective units might 
fail early on, but as time goes on and defective units get weeded out, the probability
of a unit failing decreases. 
For $k > 1$ the probability of failure increases with time. This type of failures are 
called ``wear-out'' failures, when units become more likely
to fail with time. For $k = 1$ the failure rate is constant and the distribution
is equivalent to the exponential distribution.

\subsection{Data}

The dataset we use comes from user interaction data from a major ad supported
music streaming service.
We define a user session as a period of continuous listening, demarcated
by breaks or pauses of 30 minutes or longer, i.e. a new session is started 
if a user stops or pauses the music for 30 minutes or more.
We gathered data from a random subset of users for a period of 3 months (February-April 2016),
resulting in 4,030,755 sessions.

\begin{figure}
    \centering 
    \includegraphics[width=0.47\textwidth]{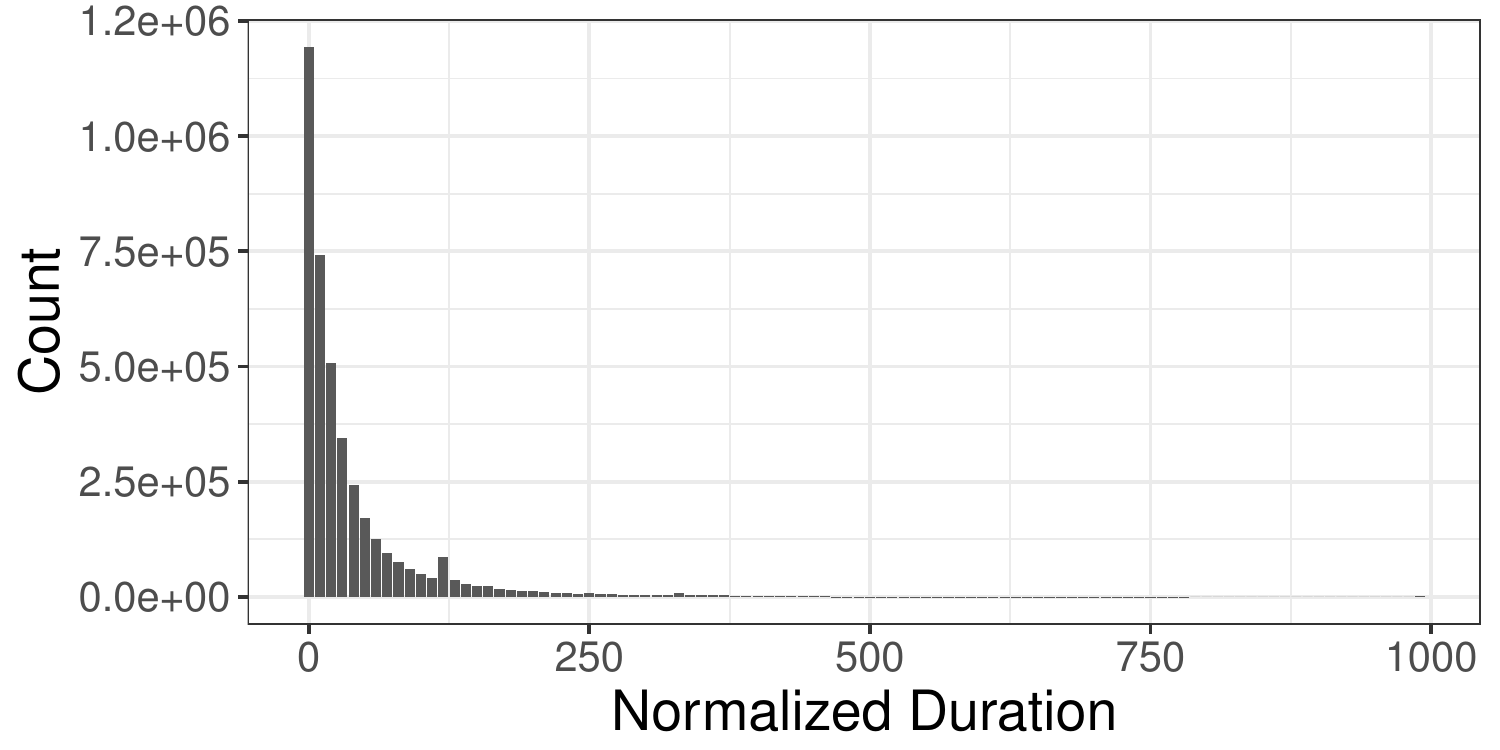}
    \caption{Histogram plot of session length. The x-axis has been normalized to the 1-1000 range.}
    \label{fig:duration-hist}
\end{figure}

In Figure \ref{fig:duration-hist} we can see a histogram for the session length data. For
confidentiality reasons the x-axis has been normalized to 1000 bins.
The distribution is highly skewed to the right, with a very small number of sessions
going all the way up to the cutoff.

\subsection{Analysis of user session length distribution}

\label{subsec:weibull-analysis}

For our analysis, we fit a Weibull distribution on the data of each user using Maximum Likelihood
Estimation with the \texttt{fitdistrplus} R package \cite{delignette2015fitdistrplus}.

\begin{figure}
	\centering
	\includegraphics[width=0.47\textwidth]{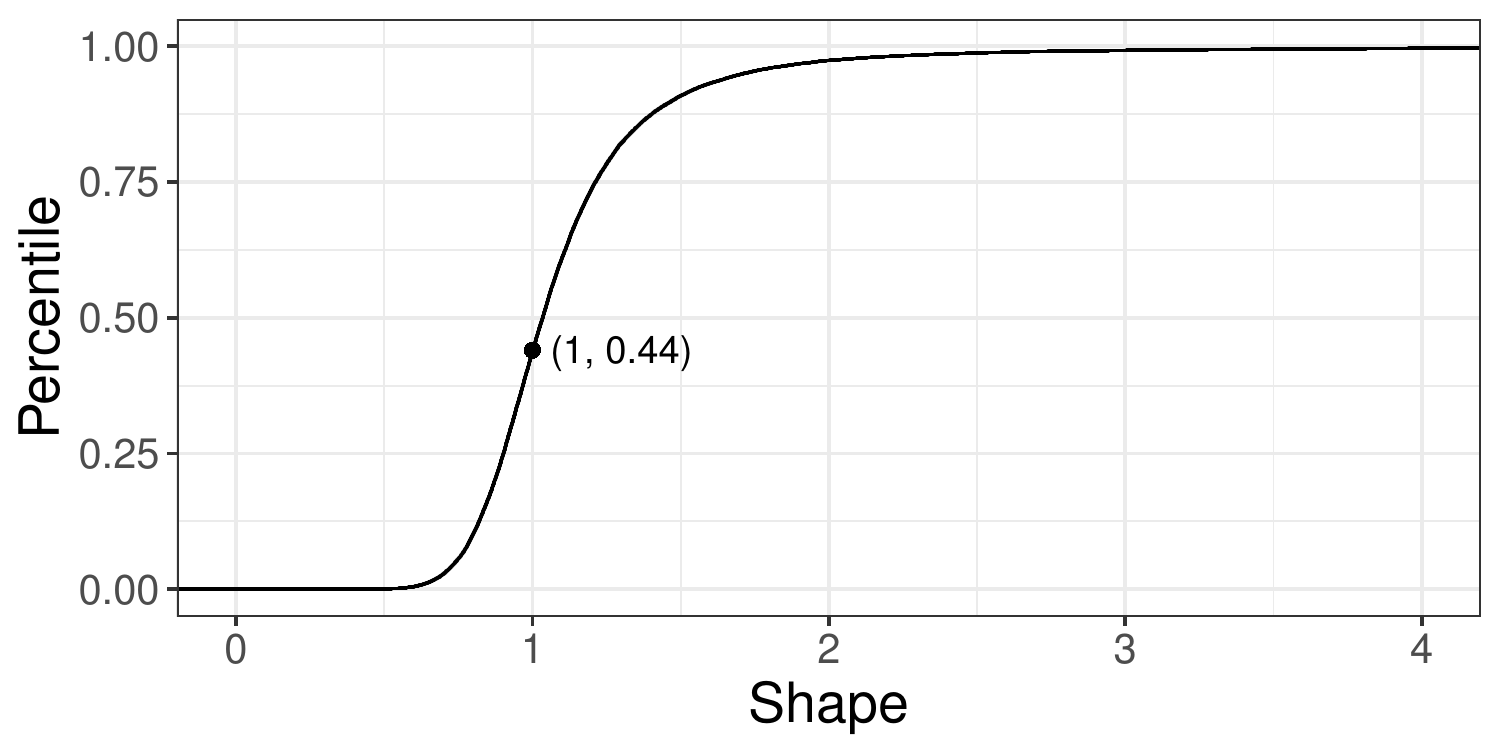}
	\caption{The empirical cumulative distribution for the shape parameter per user.
		The x axis has been truncated at $x=4$ for readability (~99.5 \% of data points shown).}
	\label{fig:shape-ecdf}
\end{figure}

In Figure \ref{fig:shape-ecdf} we can see the empirical cumulative distribution for the
shape parameter. We observe that the users in our sample are split approximately
down the middle, with 44\% of the users exhibiting Weibull distributions with $k <= 1$ and the rest $k > 1$.
Although not directly comparable, we note that for the dwell time on web sites after a search, 98.5\%
of the web sites visited have dwell time distributions with $k < 1$, exhibiting
almost exclusively the ``negative aging'' effect \cite{liu2010weibull}.

One consideration we should note here is that the variability in $k$ could also be caused by sampling
variability between users. We aim to investigate this through hypothesis testing in
an extended version of this work.

\section{Prediction}
\label{sec:prediction}

Apart from investigating the distribution of session length, ultimately we would like to
be able to predict the length of a session. To that end we gathered
features about the users and sessions, and treated the problem of predicting the
length of a session as a regression problem.

\subsection{Features}

For each of the users and sessions available in our sample we collected
a number of features.
Some features, which we call ``user-based''
are features that we assume do not change between sessions, for example
the gender of a user. Other features which we call ``contextual'' can change
every time a user starts a new session, for example the type of network or device
that a user was using when they started the session, or the length of their last session.
We provide a summary of some of these
features in Table \ref{tab:prediction-features}, separated into user-based and contextual
features.


\subsection{Model}

\label{subsec:model}

We selected gradient boosted trees (GBTs) \cite{friedman2001gbt} as our model
for a number of reasons: First, because our dataset contains missing data, the algorithm we chose had to be able to handle them explicitly, which decision trees do.

Second, the method should allow for proper modeling of non-negative data.
For such data it is possible to log-transform the dependent and use a squared error objective,
but using an objective function that is better fitted to the distribution of the dependent is
often desirable, which is a common use-case for Generalized Linear Models.
GBTs provide a flexible
optimization framework that allowed us to do just that.
To test both approaches,
we first log-transformed the dependent
and used a root mean squared error objective,
then ran the same experiments again, this time selecting the log-likelihood objective of a Gamma distribution with a log link function,
which allows for explicit modeling of right-skewed, non-negative data. In Section \ref{sec:experiments} we refer to these models
as \textit{linear} and \textit{Gamma} respectively.

Finally, we tested two versions of each model. One aggregated where a single model was created
using the data of all the users, and one per-user, where we separately trained one model
per user, using only the data originating from that user for the training and testing.
This meant that only contextual and not user-level features could be used
to train the per-user models.
This way we tested the trade-off between
the statistical power that a large dataset provides versus having personalized models.

\begin{table}
	\caption{Example user-based and contextual features used in the models.}
	\label{tab:prediction-features}
	\begin{tabular}{ll}
		\toprule
		Feature & Description\\
		\midrule
		Gender & The gender of the user \\
		Age & The age of the user \\
		Subscription Status & Whether the account is ad-supported \\
		\midrule
		Device & The device used for the session \\
		Network & The type of network used for the session \\
		Previous duration & The duration of the user's last session \\
		Absence time & Time elapsed since the last session \\
		\bottomrule
	\end{tabular}
\end{table}


\section{Experiments}

\label{sec:experiments}

The baseline model we tested against is the per-user mean session length;
that is, we calculated the mean session length in the training set for each user, and used that value
to make all predictions for each session that user had in the test set. This gave us a baseline
that is simple, but personalized to account for the differences in listening habits between
users.

Because we are focusing on direct length prediction rather than survival probability
\cite{Barbieri2016RSFclick}, thresholded duration classification \cite{lalmas2015gemini}, or
distribution parameter estimation \cite{Kim2014satisfaction, liu2010weibull}, most of the
related work models used in search and ad click scenarios are not directly applicable. Therefore
we don't include them in our comparison.

We used 10-fold cross validation, and stratified our sample per user to ensure that every user
had data points both in the train and test set of each split.

To ensure that we have enough data points per user, we only retained users
that had at least 20 sessions recorded.
The resulting dataset had 3,563,544 sessions.
Due to the size of the dataset we chose to use
the \textit{xgboost} \cite{Chen2016xgboost} variant of GBTs which is
implemented with scalability in mind, utilizing parallel, cache-aware,
and out-of-core computation to handle massive data sets. The parameters
for xgboost were selected through cross-validation on a separate validation set.


\subsection{Metrics}

We chose two evaluation metrics to measure the performance of our algorithms.
The first was the Root Mean Square Error (RMSE), which is a common choice for regression problems.
In particular we used the normalized variant of the measure (nRMSE), which is simply the RMSE scaled
by the mean value of the dependent, $\bar{y}$.

Large errors can be observed more often when the distribution
of the dependent variable is highly skewed as it is in our case (see Figure \ref{fig:duration-hist}).
Therefore, we include the Median Absolute Error (MAE) in our analysis
due to its robustness to outliers. This way
a few very large errors will not affect the metric disproportionately, compared to taking the mean.
For confidentiality we normalize all the MAE measurements by the baseline so that it has has an
error of 1, and lower measurements are better.

\subsection{Results}

We report the performance of the various approaches in Table \ref{tab:prediction-metrics}.  As mentioned before, we refer to the models using the RMSE objective as \textit{linear} to avoid confusion
with the nRMSE metric. The linear aggregated model outperforms all models in terms of Median Absolute Error, but cannot beat the baseline on nRMSE.
The aggregated model using Gamma regression has the best performance in terms of nRMSE, but has worse
MAE than its linear counterpart. We note that it's the only model that beats the baseline in
nRMSE.

The per-user linear models outperform the baseline for MAE but not for nRMSE, similarly to the aggregated linear model.
The per-user models using Gamma regression perform similarly to the linear per-user models,
indicating that the change in objective function becomes less important in small data domains.

\begin{table}
	\caption{Performance metrics for length prediction task. We report the
		mean value across the 10 CV folds, and the standard deviation in parentheses.}
	\label{tab:prediction-metrics}
	\begin{tabular}{lll}
		\toprule
		Method (\textit{Objective}) & Normalized MAE & nRMSE \\
		\midrule
		Baseline & 1 \textit{(0.001)} & 1.16 \textit{(0.005)} \\
		Aggregated (\textit{Linear}) & \textbf{0.71} \textit{(0.008)} & 1.23 \textit{(0.008)} \\
		Aggregated (\textit{Gamma}) & 0.93 \textit{(0.007)} & \textbf{1.10} \textit{(0.005)} \\
		Per-user (\textit{Linear}) & 0.83 \textit{(0.002)} & 1.29 \textit{(0.004)} \\
		Per-user (\textit{Gamma}) & 0.86 \textit{(0.001)} & 1.31 \textit{(0.003)} \\
		\bottomrule
	\end{tabular}
\end{table}

What these results indicate is that
the aggregated model using the Gamma regression objective is able to place the mean of the distribution more accurately because it
places more probability mass in the right tail of the distribution. The linear models place more of their probability mass closer
to the origin, allowing them to better capture the shorter sessions that are over-represented in the data, but as
a result miss many of the longer (outlier) sessions. This causes their mean-based metrics to suffer, while median metrics
benefit.

We also see that the per-user models mostly perform worse than their aggregated counterparts. This can be explained by the
fact that per-user models are mostly trained on few data points, and for a flexible model like GBTs, they are likely to
overfit. In this case the trade-off between having a single model trained with all the data versus
having personalized models trained on each user's data favors the aggregated model.

\section{Conclusions}
\label{sec:conclusions}
%

In this paper we perform the first, to the best of our knowledge, analysis of session
length in a mobile-focused online service.
Our analysis showed that the probability of a session to end evolves differently
for different users, with some exhibiting ``negative aging`` and others
``positive aging''.
We also used a state of the art prediction algorithm to predict the length
of a session based on user-level and contextual features, and demonstrated the differences
between using a root mean squared error objective function and a Gamma regression
objective that is more suited to right-skewed, non-negative data such as session length.


In the future we aim to demonstrate the utility of session length prediction,
by using the predictions made by our model as input to a recommender system,
or an ad scheduling algorithm.
We also plan to improve our predictive models by incorporating features
whose values evolve during a session (like interactions with the application), and keep a running
estimate for the duration of a session as it evolves.


%
\bibliographystyle{ACM-Reference-Format}
\bibliography{paper}

\end{document}